\documentstyle[12pt,amssymb,amscd]{amsart}
\newcommand{\ccc}{{\Bbb C}}

\newcommand{\zzz}{{\Bbb Z}}
\newcommand{\jj}{{\frak{g}}}
\newcommand{\hhh}{{\frak{h}}}
\newcommand{\aaa}{{\frak{A}}}
\newcommand{\iii}{{\frak{I}}}
\newcommand{\pa}{\partial}

\newcommand{\ind}{}
\newenvironment{thm}[1]{\smallskip\ind{\sf #1.}\sl}{\smallskip}

\begin{document}
\title{ A construction of N=2 and centerless N=4 superconformal fields  
via affine superalgebras}
\maketitle
\vspace{-2mm}
\begin{center}
Minoru Wakimoto
\end{center}
\vspace{2mm}
\begin{center}
{\small
 Graduate School of Mathematics, Kyushu University, Fukuoka 812-81, Japan}
\end{center}
\vspace{7mm}
\begin{center}
{\bf Abstract}
\end{center}
\begin{center}
\vspace{3mm}
\parbox{12cm}{
In this note we give a new construction of the N=2 superconformal algebra 
using currents of the affine superalgebra $\widehat{sl}(2 \vert 1)$ and 
free bosonic fields, and also the N=4 superconformal algebra without 
central charge in terms of currents of $\widehat{sl}(2 \vert 2)$ and free 
bosonic fields.} 
\end{center}
\section*{ 0. 
Introduction}

There seem to be several ways to construct the N=2 superconformal algebra 
via affine Lie algebras or affine superalgebras by tensoring with some 
fermionic or bosonic fields. In this note, first we give a simple method 
to construct it from the affine superalgebra $\widehat{sl}(2 \vert 1)$ with 
the help of two free bosonic fields and their dual fields. This construction 
is achieved by a calculus of the inner product and the structure of root 
systems. A very particular feature is that the similar construction does 
not work for $\widehat{sl}(3,\ccc)$ to get a finite-rank algebra, but only 
gives an algebra of infinite rank just like the famous Zamolodchikov's 
W-algebra.  For $\widehat{sl}(2 \vert 1)$, however, the operator product is 
well-behaved and gives a family of finite fields,  namely the N=2 
superconformal algebra. We see in this construction that the existence of 
odd simple roots with zero square length play a very important role. So our 
calculus is not described only by the terminology of root systems,  but the 
inner product is concerned.

Since the construction works so well for the simplest superalgebra 
$\widehat{sl}(2 \vert 1)$, one will naturally ask how to extend this method 
to get higher rank superconformal algebras.  It is not clear, however, 
whether superconformal algebras with fields of negative conformal weights 
are expected to be obtained in such a way associated to affine superalgebras. 
It may be expected for {\it physical} superconformal algebras which, by 
definition, do not admit fields of non-positive conformal weights and are 
completely classified by \cite{K4}. But actually the difficulty arises soon 
in the higher rank case to get a family of finite numbers of fields closed 
under the operator product.  In this note, we propose a way to construct 
the N=4 superconformal algebra, although with zero central charge, 
associated to the affine superalgebra  $\widehat{A}(1,1)$, namely the 
affinization of the finite-dimensional simple superalgebra $sl(2 \vert 2)$ 
divided by its center.  We find a family of finite fields which are 
{\it almost closed} under the operator products, and an algebra of finite 
rank is obtained by factorizing this vertex algebra by an ideal with a 
single generator.  This procedure of factorization, however, enforces us 
to retreat to the zero central charge.

In our construction, the N=2 superconformal algebra acts on the space of 
tensor procduct of an irreducible highest weight 
$\widehat{sl}(2 \vert 1)$-module and the Clifford module, and also similar
for the N=4 superconformal algebra.  We find its invariant subspaces by 
using commutants.  The analysis of their irreducibility and characters 
remains a problem.

\section*{1. Some formulas for operator products}

Throughout this paper we make use of the physicists' convention 
of the operator product
$$a(z)b(w)= \sum^{n}_{j=1} \frac{c_{j}(w)}{(z-w)^{j}}$$
to denote 
$$[a(z),b(w)] =\sum^{n}_{j=1} c_{j}(w)\frac{\pa^{j-1}_{w}}{(j-1)!}
\delta(z-w).$$ 

Let $\jj$ be a finite-dimensional simple superalgebra with an even 
non-degenerate super-invariant super-symmetric bilinear form 
$(\,\ \vert \,\ )$, and $\hat{\jj}$ be its affinization.  Here \lq\lq even" 
means, by definition, that the even part of $\jj$ is orthogonal to the odd 
part with respect to $(\,\ \vert \,\ )$ (cf. \cite{K1}).  Let $\Delta$ be 
the set of all roots of $\jj$ with respect to its Cartan subalgebra $\hhh$, 
and $\Delta^{+}$ (resp. $\Delta^{-}$) the subset of $\Delta$ consisting of 
all positive (resp. negative) roots.  Also let $\Delta_{even}$ (resp. 
$\Delta_{odd}$) denote the set of all even (resp. odd) roots.  As usual, for 
a root $\alpha$, $X_{\alpha}$ is an element from the corresponding root 
space $\jj_{\alpha}$, and $H_{\alpha}$ is the element in $\hhh$ satisfying 
$\alpha(h) = (H_{\alpha} \vert h)$ for all $h \in \hhh$.

For each $\alpha \in \Delta^{+}$, we introduce two fields 
of conformal weight $\frac{1}{2}$:
$$\psi_{\alpha}(z)=\sum_{n \in \zzz}\psi_{\alpha}(n)z^{-n-\frac{1}{2}}
=\sum_{n \in \frac{1}{2}+\zzz}\psi_{\alpha}(n+\frac{1}{2})z^{-n-1},$$
and
$$\psi^{\alpha}(z)=\sum_{n \in \zzz} \psi^{\alpha}(n)z^{-n-\frac{1}{2}}
=\sum_{n \in \frac{1}{2}+\zzz} \psi^{\alpha}(n+\frac{1}{2})z^{-n-1}.$$
These fields are fermionic or bosonic according as the associated root 
$\alpha$ is even or odd, and assumed to satisfy the (anti-)commutation 
relations
\begin{equation*}
[\psi^{\alpha}(m),\psi_{\beta}(n)]=(-1)^{p(\alpha)}[\psi_{\beta}(n),
\psi^{\alpha}(m)] =\delta_{\alpha,\beta}\delta_{m+n,0}, \tag{1.1}
\end{equation*}
where $p(\alpha)$ denotes the parity of a root $\alpha$, namely is equal 
to 0 if $\alpha$ is even and 1 if $\alpha$ is odd.
We extend $\psi^{\alpha}$ and  $\psi_{\alpha}$  to negative roots by
$\psi_{-\alpha}(z)=(-1)^{p(\alpha)}\psi^{\alpha}(z)$ and 
$\psi^{-\alpha}(z)=\psi_{\alpha}(z)$
for each positive root $\alpha$. These relations are simply written as 
\begin{equation*}
\psi_{\alpha}(z)=\epsilon(\alpha)\psi^{-\alpha}(z),  \tag{1.2a}
\end{equation*}
 and 
\begin{equation*}
\psi^{\alpha}(z)=(-1)^{p(\alpha)}\epsilon(\alpha)\psi_{-\alpha}(z), 
 \tag{1.2b}
\end{equation*}
for all roots $\alpha$, where 
\begin{equation*}
\epsilon(\alpha) :=
\begin{cases}
 1   &  \qquad \text{if}\,\  \alpha \,\  \text{is positive} \\
(-1)^{p(\alpha)}  & \qquad \text{if} \,\ \alpha \,\ \text{is negative}.
\end{cases}
\end{equation*}

For a positive root $\alpha$,  the operators $\psi^{\alpha}(m)$ with 
$m < 0$  (resp. $m \ge 0$) and $\psi_{\alpha}(n)$ with $n \le 0$ 
(resp. $n > 0$) are assumed to be creation (resp. annihilation) 
operators. Using the inner product
\begin{equation*}
(\psi^{\alpha} \vert \psi^{\beta}) = (\psi_{\alpha} \vert \psi_{\beta})
:= \epsilon(\alpha) \delta_{\alpha+\beta,0},  \tag{1.4a}
\end{equation*}
or equivalently
\begin{equation*}
(\psi^{\alpha} \vert \psi_{\beta}) 
= (-1)^{p(\alpha)}(\psi_{\alpha} \vert \psi^{\beta})
:= \delta_{\alpha,\beta},  \tag{1.4b}
\end{equation*}
the operator products of these fields are simply written as the following 
formula:
\begin{equation*}
\varphi(z) \psi(w) = \frac{(\varphi \vert \psi)}{z-w},  \tag{1.5}
\end{equation*}
where $\varphi(z)$ and $\psi(z)$ are elements from $\psi^{\alpha}(z)$'s 
and $\psi_{\alpha}(z)$'s.

For each element $X \in \jj$, the field 
$$X(z) := \sum_{n \in Z} X(n) z^{-n-1}$$
is called a {\it current}, where $X(n)$'s, $n \in \zzz$, are elements in the
affinization $\widehat{\jj}$ of $\jj$, satisfying the (anti-)commutation 
relation
\begin{equation*}
 [X(m), Y(n)] = [X,Y](m+n) + m(X \vert Y) \delta_{m+n,0}K, \tag{1.6a}
\end{equation*}
where $K$ is the canonical central element in $\widehat{\jj}$.

In terms of the operator product, the formula (1.6a) is written as
\begin{equation*}
X(z)Y(w) = \frac{(X \vert Y)K}{(z-w)^2} + \frac{[X,Y](w)}{z-w}.  \tag{1.6b}
\end{equation*}
In this note, 
$a(z)$, $b(z)$, $c(z), \cdots$, $X(z)$, $Y(z), \cdots$ always stand for 
currents, and $\psi(z)$, $\varphi(z), \cdots$  free fields with product 
(1.5), and $p(\,\ )$ denotes the parity of a field. We note that
\begin{equation*}
p(\psi^{\alpha}) = p(\psi_{\alpha}) = p({\alpha}) +1 \qquad  \text{mod}\,\ 
2\zzz.    \tag{1.7}
\end{equation*}

A very important tool in our operator calculus is the Wick theorem in 
particular for non-commutative fields, a clear explanation on which is given, 
e.g., in Section 3.3 of \cite{K3}. Here is a list of formulas obtained from 
the non-commutative Wick theorem, which will be used in this paper.
{\allowdisplaybreaks %
\begin{equation*}
\begin{split}
  :a(z &)b(z)::c(w):   
= \frac{-([a,b] \vert c)K}{(z-w)^3} +(b \vert c)K 
\left\{\frac{a(w)}{(z-w)^2} + \frac{\pa a(w)}{z-w} \right\}   \\
&+(-1)^{p(b)p(c)}(a \vert c)K \left\{\frac{b(w)}{(z-w)^2} 
+ \frac{\pa b(w)}{z-w} \right\} 
+(-1)^{p(a)p(b)} \frac{[b,[a, c]](w)}{(z-w)^2}  \\
&+ \frac{1}{z-w} \left\{:a(w)[b, c](w): + (-1)^{p(a)p(b)}:b(w)[a,c](w): 
\right\},
\end{split}       \tag{1.8a}
\end{equation*}
}
{\allowdisplaybreaks %
\begin{equation*}
\begin{split}
:a(&z)::b(w)c(w): = \frac{(a \vert [b,c])K}{(z-w)^3}   \\
&+\frac{1}{(z-w)^2}\left\{ (a \vert b)K \cdot c(w)
+(-1)^{p(a)p(b)}(a \vert c)K \cdot b(w) +[[a,b],c](w) \right\} \\
&+ \frac{1}{z-w} \left\{:[a,b](w)c(w): + (-1)^{p(a)p(b)}:b(w)[a,c](w): 
\right\},
\end{split}      \tag{1.8b}
\end{equation*}
}
{\allowdisplaybreaks %
\begin{equation*}
\begin{split}
(-1&)^{p(\varphi)p(Y)}:X(z)\varphi(z)::Y(w)\psi(w): 
= (X \vert Y)(\varphi \vert \psi)\frac{K}{(z-w)^3}   \\
&+(\varphi \vert \psi)\frac{[X,Y](w)}{(z-w)^2}
+(X \vert Y)K \left\{\frac{:\varphi(w)\psi(w):}{(z-w)^2}
+\frac{:\pa \varphi(w)\psi(w):}{z-w}\right\}  \\
&+\frac{1}{z-w} \left\{[x,y](w): \varphi(w) \psi(w):
+ (\varphi \vert \psi) :X(w)Y(w): \right\},
\end{split}    \tag{1.9}
\end{equation*}
}
{\allowdisplaybreaks %
\begin{equation*}
\begin{split}
(-1&)^{p(X)p(\varphi)} :X(z)\varphi(z)::Y(w)Z(w):   \\
&= ([X,Y] \vert Z)K
\left\{\frac{\varphi(w)}{(z-w)^3}+\frac{\pa \varphi(w)}{(z-w)^2}
+\frac{\frac{1}{2} \pa^2 \varphi(w)}{z-w} \right\} \\
&+(X \vert Y)K
\left\{\frac{\varphi(w)Z(w)}{(z-w)^2}+\frac{\pa \varphi(w)Z(w)}
{z-w}\right\} \\
&+(-1)^{p(X)p(Y)}(X \vert Z)K
\left\{\frac{\varphi(w)Y(w)}{(z-w)^2}+\frac{\pa \varphi(w)Y(w)}
{z-w}\right\}   \\
&+\frac{\varphi(w)[[X,Y],Z](w)}{(z-w)^2}
+\frac{\pa \varphi(w)[[X,Y],Z](w)}{z-w}  \\
&+\frac{1}{z-w}\varphi(w):[X,Y](w)Z(w): 
+\frac{(-1)^{p(X)p(Y)}}{z-w} \varphi(w):Y(w)[X,Z](w):,
\end{split}        \tag{1.10}
\end{equation*}
}
{\allowdisplaybreaks %
\begin{equation*}
\begin{split}
& (-1)^{p(Y)(p(\varphi_1)+p(\varphi_2))}
:X(z)\varphi_1(z)\varphi_2(z)::Y(w)\psi(w):   \\
&=(X \vert Y)K(\varphi_2 \vert \psi) \left\{\frac{\varphi_1(w)}{(z-w)^3}
+\frac{2 \pa \varphi_1(w)}{(z-w)^2}
+\frac{\frac{3}{2} \pa^2 \varphi_1(w)}{z-w} \right\}  \\
&+(-1)^{p(\varphi_2)p(\psi)}(X \vert Y)K(\varphi_1 \vert \psi)
\left\{\frac{\varphi_2(w)}{(z-w)^3}+\frac{2 \pa \varphi_2(w)}{(z-w)^2}
+\frac{\frac{3}{2} \pa^2 \varphi_2(w)}{z-w} \right\}   \\
&+(X \vert Y)K \left\{\frac{:\varphi_1(w) \varphi_2(w) \psi(w):}{(z-w)^2}
+\frac{:\pa (\varphi_1(w) \varphi_2(w)) \psi(w):}{z-w}  \right\}  \\
&+(\varphi_2 \vert \psi) \left\{\frac{[X,Y](w)\varphi_1(w)}{(z-w)^2}
+\frac{[X,Y](w) \pa \varphi_1(w)}{z-w}
+\frac{:X(w)Y(w): \varphi_1(w)}{z-w} \right\}   \\
&+(-1)^{p(\varphi_2)p(\psi)}(\varphi_1 \vert \psi)
\left\{\frac{[X,Y](w)\varphi_2(w)}{(z-w)^2}
+\frac{[X,Y](w) \pa \varphi_2(w)}{z-w}
+\frac{:X(w)Y(w): \varphi_2(w)}{z-w} \right\}   \\
&+\frac{[X,Y](w):\varphi_1(w) \varphi_2(w) \psi(w):}{z-w}.
\end{split}        \tag{1.11}
\end{equation*}
}

In particular from (1.8) and (1.9) one deduces the following:

\begin{thm}{Lemma 1.1} Let $\alpha, \beta \in \Delta$ \,\ such that
$\alpha + \beta \ne 0$, then 
\begin{equation*}
:X_{\alpha}(z) \psi^{\alpha}(z)::X_{\beta}(w) \psi^{\beta}(w):
= \frac{(-1)^{(p(\alpha)+1)p(\beta)}}{z-w}[X_{\alpha}, X_{\beta}](w)
:\psi^{\alpha}(w)\psi^{\beta}(w):. 
\end{equation*}
\end{thm}

\begin{thm}{Lemma 1.2}
Let $H \in \hhh$ and $X_{\alpha} \in \jj_{\alpha}$
and $X_{-\alpha} \in \jj_{-\alpha}$ such that 
$[X_{\alpha}, X_{-\alpha}]= H_{\alpha}$;  then 
\begin{enumerate}
\renewcommand{\labelenumi}{\arabic{enumi})}
\item 
$$ H(z):X_{\alpha}(w)X_{-\alpha}(w): = \frac{(H_{\alpha} \vert H)K}{(z-w)^3}
+\alpha(H) \frac{H_{\alpha}(w)}{(z-w)^2}, $$
\item

$$ :X_{\alpha}(z)X_{-\alpha}(z):H(w) = \frac{-(H_{\alpha} \vert H)K}{(z-w)^3}
+\alpha(H) \left\{\frac{H_{\alpha}(w)}{(z-w)^2} 
+\frac{\pa H_{\alpha}(w)}{z-w} \right\}, $$
\item 
\begin{align*}
& :X_{\alpha}(z) \psi^{\alpha}(z)::X_{-\alpha}(w) 
\psi^{-\alpha}(w): \\
&= \epsilon(\alpha)\frac{(X_{\alpha} \vert X_{-\alpha})K}{(z-w)^3} 
+\frac{\epsilon(\alpha)}{(z-w)^2}\left\{(X_{\alpha} \vert X_{-\alpha})K
:\psi^{\alpha}(w)\psi_{\alpha}(w): + H_{\alpha}(w)\right\}  \\
&+\frac{\epsilon(\alpha)}{z-w} \{:X_{\alpha}(w)X_{-\alpha}(w):
+ H_{\alpha}(w):\psi^{\alpha}(w)\psi_{\alpha}(w): \\
&+(X_{\alpha} \vert X_{-\alpha})K :\pa \psi^{\alpha}(w)
\psi_{\alpha}(w): \}.
\end{align*}
\end{enumerate}
\end{thm}

\section*{2. The N=2 superconformal algebra associated to 
$\widehat{sl}(2 \vert 1)$}

In this section we start our discussion under a general situation, where 
$\jj$ is a simple Lie superalgebra with an even non-degenerate 
super-invariant super-symmetric bilinear form $(\,\ \vert \,\ )$. 
One can choose $H_{\alpha}$ and $X_{\alpha}$ such that
\begin{equation*}
(X_{\alpha} \vert X_{-\alpha}) = (-1)^{p(\alpha)} \epsilon(\alpha), \tag{2.1}
\end{equation*}
\begin{equation*}
H_{\alpha} = (-1)^{p(\alpha)} \epsilon(\alpha) \alpha. \tag{2.2a}
\end{equation*}
Note that
\begin{equation*}
H_{-\alpha} = -(-1)^{p(\alpha)} H_{\alpha}. \tag{2.2b}
\end{equation*}

Given a pair ($\alpha$, $\beta$) of roots such that 
\begin{equation*}
\text{ (the linear span of} \,\ \alpha \,\ \text{and} \,\ \beta)
 \cap \Delta = \{ \pm \alpha,  \pm \beta, \pm (\alpha + \beta) \} ,
 \tag{2.3}
\end{equation*}
we introduce the following fields:
{\allowdisplaybreaks %
\begin{align*}
G_{+}(z) &:= X_{\alpha}(z)\psi^{\alpha}(z)+ X_{-\beta}(z)\psi^{-\beta}(z), \\
G_{-}(z) &:= X_{\beta}(z)\psi^{\beta}(z)+ X_{-\alpha}(z)\psi^{-\alpha}(z), \\
A(z) &:= K \{(-1)^{p(\alpha)}\epsilon(\alpha):\psi^{\alpha}(z)
\psi_{\alpha}(z):
- (-1)^{p(\beta)}\epsilon(\beta):\psi^{\beta}(z) \psi_{\beta}(z): \}  \\
&+H_{\alpha}(z)-H_{\beta}(z), \\
B(z) &:= K \{(-1)^{p(\alpha)}\epsilon(\alpha):\pa \psi^{\alpha}(z) 
\psi_{\alpha}(z):
- (-1)^{p(\beta)}\epsilon(\beta):\psi^{\beta}(z) \pa \psi_{\beta}(z): \} \\
&+:X_{\alpha}(z)X_{-\alpha}(z): + :X_{\beta}(z)X_{- \beta}(z): 
- \pa H_{\beta}(z)  \\
&+ H_{\alpha}(z):\psi^{\alpha}(z) \psi_{\alpha}(z):  
+ H_{\beta}(z):\psi^{\beta}(z) \psi_{\beta}(z):  \\
&+(-1)^{p(\alpha)p(\beta)}c_{\alpha, \beta}  \{
(-1)^{p(\beta)}:X_{\alpha+\beta}(z)\psi^{\alpha}(z) \psi^{\beta}(z): \\
&-(-1)^{p(\alpha)}\epsilon(\alpha)
\epsilon(\beta):X_{-\alpha-\beta}(z)\psi_{\alpha}(z) \psi_{\beta}(z): \}.
\end{align*}
}

In the actual calculation of operator products of fields, one needs the 
following relations among structure constants, which are deduced from 
the Jacobi identity:
\begin{equation*}
\begin{split}
c_{-\alpha, -\beta} &= - (-1)^{p(\alpha+\beta)}c_{\alpha, \beta},  \\
c_{\alpha,-\alpha -\beta} &= - (-1)^{(p(\alpha)+1)p(\beta)}
   \epsilon(\alpha)c_{\alpha, \beta},  \\
c_{\beta, -\alpha -\beta} &= (-1)^{p(\alpha)} \epsilon(\beta)
c_{\alpha, \beta}, \\
c_{-\alpha, \alpha +\beta} &=  (-1)^{p(\alpha)p(\beta)}
   \epsilon(\alpha)c_{\alpha, \beta},  \\
c_{-\beta, \alpha +\beta} &=  - \epsilon(\beta)c_{\alpha, \beta}.  
\end{split}     \tag{2.4}
\end{equation*}

Then, by Lemmas 1.1 and 1.2, one easily has the following:

\begin{thm}{Lemma 2.1}
\begin{enumerate}
\renewcommand{\labelenumi}{\arabic{enumi})}
\item \qquad
$$ G_{+}(z)G_{+}(w) = G_{-}(z)G_{-}(w)=0,   $$
\item 
\begin{align*}
&G_{+}(z)G_{-}(w) \\
&=\{(-1)^{p(\alpha)}\epsilon(\alpha)+(-1)^{p(\beta)}\epsilon(\beta) \}
\frac{K}{(z-w)^3} +\frac{A(w)}{(z-w)^2}+\frac{B(w)}{z-w}.
\end{align*}
\end{enumerate}
\end{thm}

And the following is shown by a simple calculation using (1.8),(1.9) and 
(1.10):  \\

(2.5a) \qquad  $A(z) :X_{\alpha}(w)\psi^{\alpha}(w) : $
\begin{equation*}
=\{\epsilon(\alpha)K 
+ (-1)^{p(\alpha)}\epsilon(\alpha)(\alpha \vert \alpha) 
- (-1)^{p(\beta)}\epsilon(\beta)(\alpha \vert \beta) \}
\frac{X_{\alpha}(w) \psi^{\alpha}(w)}{z-w},
\end{equation*}
(2.5b) \qquad  $A(z) :X_{-\beta}(w)\psi^{-\beta}(w) : $
\begin{equation*}
=\{\epsilon(\beta)K 
- (-1)^{p(\alpha)}\epsilon(\alpha)(\alpha \vert \beta) 
+ (-1)^{p(\beta)}\epsilon(\beta)(\beta \vert \beta) \}
\frac{X_{-\beta}(w) \psi^{-\beta}(w)}{z-w},
\end{equation*}
(2.6a) \qquad  $A(z) :X_{\beta}(w)\psi^{\beta}(w) : $
\begin{equation*}
=\{-\epsilon(\beta)K 
+ (-1)^{p(\alpha)}\epsilon(\alpha)(\alpha \vert \beta) 
- (-1)^{p(\beta)}\epsilon(\beta)(\beta \vert \beta) \}
\frac{X_{\beta}(w) \psi^{\beta}(w)}{z-w},
\end{equation*}
(2.6b) \qquad  $A(z) :X_{-\alpha}(w)\psi^{-\alpha}(w) : $
\begin{equation*}
=\{-\epsilon(\alpha)K 
- (-1)^{p(\alpha)}\epsilon(\alpha)(\alpha \vert \alpha) 
+ (-1)^{p(\beta)}\epsilon(\beta)(\alpha \vert \beta) \}
\frac{X_{-\alpha}(w) \psi^{-\alpha}(w)}{z-w}.
\end{equation*}

  From these, one has the following:

\begin{thm}{Lemma 2.2}
Assume that 
$(\alpha \vert \alpha) = (\beta \vert \beta) =0$, $(\alpha \vert \beta) = 1$,
and $\epsilon(\alpha) = \epsilon(\beta) =1$.  Then 
\begin{enumerate}
\renewcommand{\labelenumi}{\arabic{enumi})}
\item  $$A(z)G_{+}(w) = \frac{K+1}{z-w}G_{+}(w),$$
\item  $$ A(z)G_{-}(w) = \frac{-(K+1)}{z-w}G_{-}(w).$$
\end{enumerate}
\end{thm}

We now proceed to compute the operator product of  $B(z)$ with $G_{\pm}(z)$.
After a long and tiresome calculation using (1.9), (1.10) and (1.11),
one arrives at the following:  \\

(2.7a) \qquad  $B(z) :X_{\alpha}(w)\psi^{\alpha}(w) : $
{\allowdisplaybreaks %
\begin{align*}
&=\epsilon(\alpha)\{K+(\alpha \vert \alpha) \} \left\{ 
\frac{X_{\alpha}(w)\psi^{\alpha}(w)}{(z-w)^2}
+ \frac{\pa (X_{\alpha}(w)\psi^{\alpha}(w))}{z-w}  \right\}   \\
&+(-1)^{(p(\alpha)+1)p(\beta)} \epsilon(\beta) \left\{
\frac{X_{\alpha}(w)\psi^{\alpha}(w)}{(z-w)^2}
+ \frac{\pa X_{\alpha}(w)\psi^{\alpha}(w)}{z-w}  \right\}  \\
&+(-1)^{p(\beta)} \epsilon(\beta) ( \alpha \vert \beta)
\frac{X_{\alpha}(w)\psi^{\alpha}(w)}{(z-w)^2}           \\
&+(-1)^{p(\alpha)(p(\beta)+1)} \left\{
\frac{X_{-\beta}(w)\psi^{-\beta}(w)}{(z-w)^2}
+ \frac{\pa (X_{-\beta}(w)\psi^{-\beta}(w))}{z-w}  \right\}  \\
&+(-1)^{p(\alpha)+p(\beta)}\epsilon(\alpha)\epsilon(\beta) 
\frac{c_{\alpha, \beta}}{z-w}
:X_{\alpha}(w)X_{- \alpha- \beta}(w)\psi_{\beta}(w):  \\
&-(-1)^{p(\alpha)+p(\beta)}
\frac{c_{\alpha, \beta}}{z-w}
:X_{\alpha+ \beta}(w)X_{- \beta}(w)\psi^{\alpha}(w):   \\
&+(-1)^{p(\alpha)}\epsilon(\alpha) \frac{(\alpha \vert \alpha)}{z-w}
:X_{\alpha}(w)\psi^{\alpha}(w)\psi_{\alpha}(w)\psi^{\alpha}(w):  \\
&+(-1)^{p(\beta)}\epsilon(\beta) \frac{(\alpha \vert \beta)}{z-w}
:X_{\alpha}(w)\psi^{\alpha}(w)\psi^{\beta}(w)\psi_{\beta}(w):  \\
&+(-1)^{p(\alpha)p(\beta)}\frac{\epsilon(\beta)}{z-w}
:X_{-\beta}(w)\psi_{\beta}(w)\psi_{\alpha}(w)\psi^{\alpha}(w):,
\end{align*}
}
(2.7b) \qquad $B(z):X_{-\beta}(w)\psi^{-\beta}(w):$
{\allowdisplaybreaks %
\begin{align*}
&=\epsilon(\beta)\{K+(\beta \vert \beta) \} \left\{ 
\frac{X_{-\beta}(w)\psi^{-\beta}(w)}{(z-w)^2}
+ \frac{\pa (X_{-\beta}(w)\psi^{-\beta}(w))}{z-w}  \right\}   \\
&+(-1)^{(p(\alpha)+1)p(\beta)} \left\{
\frac{X_{\alpha}(w)\psi^{\alpha}(w)}{(z-w)^2}
+ \frac{X_{\alpha}(w) \pa \psi^{\alpha}(w)}{z-w}  \right\}   \\
&+\epsilon(\beta)(\beta \vert \beta) \{1-(-1)^{p(\beta)} \}
\frac{X_{-\beta}(w)\psi^{-\beta}(w)}{(z-w)^2}     \\
&-(-1)^{p(\alpha)+p(\beta)}\epsilon(\beta) 
\frac{c_{\alpha, \beta}}{z-w}
:X_{\alpha}(w)X_{- \alpha- \beta}(w)\psi_{\beta}(w):  \\
&+\epsilon(\beta)\frac{c_{\alpha, \beta}}{z-w}
:X_{\alpha+ \beta}(w)X_{- \beta}(w)\psi^{\alpha}(w):   \\
&-(-1)^{p(\beta)} \frac{(\beta \vert \beta)}{z-w}
:X_{-\beta}(w)\psi^{\beta}(w)\psi_{\beta}(w)\psi_{\beta}(w):   \\
&-(-1)^{p(\alpha)}\epsilon(\alpha)\epsilon(\beta) 
\frac{(\alpha \vert \beta)}{z-w}
:X_{-\beta}(w)\psi_{\beta}(w)\psi^{\alpha}(w)\psi_{\alpha}(w):    \\
&+(-1)^{(p(\alpha)+1)p(\beta)}
\frac{1}{z-w}
:X_{\alpha}(w)\psi^{\alpha}(w)\psi^{\beta}(w)\psi_{\beta}(w):,
\end{align*}
}
(2.8a) \qquad $B(z):X_{\beta}(w)\psi^{\beta}(w): $
{\allowdisplaybreaks %
\begin{align*}
&=\epsilon(\beta) \{ K+(\beta \vert \beta)\} \left\{ 
\frac{X_{\beta}(w)\psi^{\beta}(w)}{(z-w)^2}
+ \frac{\pa (X_{\beta}(w)\psi^{\beta}(w))}{z-w}  \right\}    \\
&+(-1)^{p(\alpha)(p(\beta)+1)} \epsilon(\alpha) \left\{
\frac{X_{\beta}(w)\psi^{\beta}(w)}{(z-w)^2}
+ \frac{\pa X_{\beta}(w)\psi^{\beta}(w)}{z-w}  \right\}       \\
&+\epsilon(\beta)\{K+ (-1)^{p(\beta)}( \beta \vert \beta) \}
\frac{X_{\beta}(w)\psi^{\beta}(w)}{(z-w)^2}  \\
&+(-1)^{(p(\alpha)+1)p(\beta)} \left\{
\frac{X_{-\alpha}(w)\psi^{-\alpha}(w)}{(z-w)^2}
+ \frac{\pa (X_{-\alpha}(w)\psi^{-\alpha}(w))}{z-w}  \right\}  \\
&+(-1)^{(p(\alpha)+1)(p(\beta)+1)}\epsilon(\alpha)\epsilon(\beta) 
\frac{c_{\alpha, \beta}}{z-w}
:X_{\beta}(w)X_{- \alpha- \beta}(w)\psi_{\alpha}(w):         \\
&-(-1)^{(p(\alpha)+1)(p(\beta)+1)}
\frac{c_{\alpha, \beta}}{z-w}
:X_{\alpha+ \beta}(w)X_{- \alpha}(w)\psi^{\beta}(w):  \\
&+(-1)^{p(\alpha)}\epsilon(\alpha) \frac{(\alpha \vert \beta)}{z-w}
:X_{\beta}(w)\psi^{\beta}(w)\psi^{\alpha}(w)\psi_{\alpha}(w):  \\
&+(-1)^{p(\beta)}\epsilon(\beta) \frac{(\beta \vert \beta)}{z-w}
:X_{\beta}(w)\psi^{\beta}(w)\psi_{\beta}(w)\psi^{\beta}(w):     \\
&-(-1)^{(p(\alpha)+1)p(\beta)}\frac{\epsilon(\beta)}{z-w}
:X_{-\alpha}(w)\psi_{\alpha}(w)\psi^{\beta}(w)\psi_{\beta}(w):,
\end{align*}
}
(2.8b)  \qquad $B(z):X_{-\alpha}(w)\psi^{-\alpha}(w): $
{\allowdisplaybreaks %
\begin{align*}
&=\epsilon(\alpha)\{K+(\alpha \vert \alpha) \} \left\{ 
\frac{X_{-\alpha}(w)\psi^{-\alpha}(w)}{(z-w)^2}
+ \frac{\pa (X_{-\alpha}(w)\psi^{-\alpha}(w))}{z-w}  \right\}  \\
&+(-1)^{p(\alpha)(p(\beta)+1)} \left\{
\frac{X_{\beta}(w)\psi^{\beta}(w)}{(z-w)^2}
+ \frac{X_{\beta}(w) \pa \psi^{\beta}(w)}{z-w}  \right\}  \\
&+\{ \epsilon(\alpha)(K+(\alpha \vert \alpha))-
(-1)^{p(\beta)} \epsilon(\beta) ( \alpha \vert \beta) \}
\frac{X_{-\alpha}(w)\psi^{-\alpha}(w)}{(z-w)^2}            \\
&-(-1)^{(p(\alpha)+1)(p(\beta)+1)}\epsilon(\alpha) 
\frac{c_{\alpha, \beta}}{z-w}
:X_{\beta}(w)X_{- \alpha- \beta}(w)\psi_{\alpha}(w):            \\
&-(-1)^{p(\alpha)p(\beta)}\epsilon(\alpha)
\frac{c_{\alpha, \beta}}{z-w}
:X_{\alpha+ \beta}(w)X_{- \alpha}(w)\psi^{\beta}(w):             \\
&-(-1)^{p(\alpha)} \frac{(\alpha \vert \alpha)}{z-w}
:X_{-\alpha}(w)\psi^{\alpha}(w)\psi_{\alpha}(w)\psi_{\alpha}(w): \\
&-(-1)^{p(\beta)}\epsilon(\alpha)\epsilon(\beta) \frac{(\alpha \vert \beta)}
{z-w} :X_{-\alpha}(w)\psi_{\alpha}(w)\psi^{\beta}(w)\psi_{\beta}(w):  \\
&+(-1)^{p(\alpha)(p(\beta)+1)}\frac{1}{z-w}
:X_{\beta}(w)\psi^{\beta}(w)\psi^{\alpha}(w)\psi_{\alpha}(w):.   
\end{align*}
}

Looking at the above formulas, one sees that all of the extra terms cancel 
out and disappear if  the conditions
$(\alpha \vert \alpha) = (\beta \vert \beta) =0$, $(\alpha \vert \beta) = 1$,
and $\epsilon(\alpha) = \epsilon(\beta) =1$ and $c_{\alpha, \beta} =1$ are
satisfied, and has the following:

\begin{thm}{Lemma 2.3}
Assume that 
$(\alpha \vert \alpha) = (\beta \vert \beta) =0$, $(\alpha \vert \beta) = 1$,
and $\epsilon(\alpha) = \epsilon(\beta) =1$ and $c_{\alpha, \beta} =1$. 
Then 
\begin{enumerate}
\renewcommand{\labelenumi}{\arabic{enumi})}
\item  $$B(z) G_{+}(w) = (K+1) \left\{ \frac{G_{+}(w)}{(z-w)^2}
+ \frac{\pa G_{+}(w)}{z-w} \right\},$$
\item  $$B(z) G_{-}(w) = (K+1) \left\{ \frac{2G_{-}(w)}{(z-w)^2}
+ \frac{\pa G_{-}(w)}{z-w} \right\}.$$
\end{enumerate}
\end{thm}

The other products $A(z)A(w)$,  $A(z)B(w)$  and $B(z)B(w)$  are 
obtained from the above Lemmas by using the Borcherds-Jacobi identity 
(see, e.g.,  the formula (4.6.7) in \cite{K3}):
\begin{equation*}
[a_{(m)}, b_{(n)}]= \sum_{j \ge 0}\binom{m}{j}(a_{(j)}b)_{(m+n-j)}, 
\tag{2.9}
\end{equation*}
for mutually local fields  $a(z)  = \sum_{n \in \zzz}a_{(n)}z^{-n-1}$  and
$b(z)  = \sum_{n \in \zzz}b_{(n)}z^{-n-1}$.
Actually the calculation goes as follows.  Put 
$$A(z)  = \sum_{n \in \zzz}A_{(n)}z^{-n-1},$$
$$B(z)  = \sum_{n \in \zzz}B_{(n)}z^{-n-1},$$ and
$$G_{\pm}(z)  = \sum_{n \in \zzz}G_{\pm (n)}z^{-n-1}.$$
Then the above lemmas, together with skew-symmetry, give the following:
\begin{equation*}
G_{+(n)}G_{-}  =
\begin{cases}
   B     & \qquad \text{if} \,\ n=0,    \\
   A     & \qquad \text{if} \,\ n=1,    \\
   -2K     & \qquad \text{if} \,\ n=2,  \\
   0     & \qquad \text{if} \,\ n \ge 3,
\end{cases}
\end{equation*}
\begin{equation*}
G_{-(n)}G_{+}  =
\begin{cases}
   B- \pa A     & \qquad \text{if} \,\ n=0,    \\
   -A     & \qquad \text{if} \,\ n=1,          \\
   -2K     & \qquad \text{if} \,\ n=2,         \\
   0     & \qquad \text{if} \,\ n \ge 3,
\end{cases}
\end{equation*}
\begin{equation*}
A_{(n)}G_{\pm}  = -G_{\pm (n)}A  =
\begin{cases}
   \pm (K+1)G_{\pm}     & \qquad \text{if} \,\ n=0,  \\
   0     & \qquad \text{if} \,\ n \ge 1,
\end{cases}
\end{equation*}
\begin{equation*}
B_{(n)}G_{+}  =
\begin{cases}
   (K+1)\pa G_{+}     & \qquad \text{if} \,\ n=0,  \\
   (K+1) G_{+}     & \qquad \text{if} \,\ n=1,    \\
   0     & \qquad \text{if} \,\ n \ge 2,
\end{cases}
\end{equation*}
\begin{equation*}
G_{+(n)}B  =
\begin{cases}
   0     & \qquad \text{if} \,\ n=0 \,\ \text{or} \,\ n \ge 2,  \\
   (K+1) G_{+}     & \qquad \text{if} \,\ n=1,
\end{cases}
\end{equation*}
\begin{equation*}
B_{(n)}G_{-}  = G_{-(n)}B  =
\begin{cases}
   (K+1)\pa G_{-}     & \qquad \text{if} \,\ n=0,   \\
   2(K+1) G_{-}     & \qquad \text{if} \,\ n=1,     \\
   0     & \qquad \text{if} \,\ n \ge 2.
\end{cases}
\end{equation*}
So, by (2.9) applied to $a=G_{+}$ and $b=G_{-}$,  one has
\begin{align*}
[G_{+(0)}, G_{-(n)}] &= B_{(n)},       \\
[G_{+(1)}, G_{-(n)}] &= B_{(n+1)}+A_{(n)},
\end{align*}
and, using these, can compute $A_{(n)}A$, $A_{(n)}B$, $B_{(n)}A$ and 
$B_{(n)}B$;  e.g., 
\begin{equation*}
B_{(n)}A = G_{+(0)}G_{-(n)}A + G_{-(n)}G_{+(0)}A  =
\begin{cases}
(K+1) \pa A   & \qquad  \text{if} \,\ n=0,    \\
(K+1)A   & \qquad  \text{if} \,\ n=1,         \\
2K(K+1)  & \qquad  \text{if} \,\ n=2,         \\
0   & \qquad  \text{if} \,\ n \ge 3.
\end{cases}
\end{equation*}

Translating into the terminology of operator products, 
one has the following:

\begin{thm}{Lemma 2.4}
{\allowdisplaybreaks %
\begin{align*}
A(z)A(w)  &= \frac{-4K(K+1)}{(z-w)^2},   \\
A(z)B(w)  &= (K+1) \left\{ \frac{-2K}{(z-w)^3} +\frac{A(w)}
{(z-w)^2} \right\},   \\
B(z)A(w)  &= (K+1) \left\{ \frac{2K}{(z-w)^3} +\frac{A(w)}{(z-w)^2} 
+\frac{\pa A(w)}{z-w} \right\},   \\
B(z)B(w)  &= (K+1) \left\{ \frac{2B(w)}{(z-w)^2} +\frac{\pa B(w)}
{z-w} \right\}.
\end{align*}
}
\end{thm}

Summing up the above, one obtains the following :

\begin{thm}{Theorem 2.1}
Let $\{ \alpha_{1}, \alpha_{2} \}$ 
(resp. $\{ \alpha^{\vee}_{1}, \alpha^{\vee}_{2} \}$) 
be the set of simple roots  (resp. simple coroots) of the Lie superalgebra 
$sl(2 \vert 1)$  such that
\begin{equation*}
\left( \langle \alpha^{\vee}_{i}, \alpha_{j}\rangle \right)_{i, j = 1,2} =
\left( ( \alpha^{\vee}_{i} \vert \alpha^{\vee}_{j}) \right)_{i, j = 1,2} =
\begin{pmatrix}
   0  &  1  \\
   1  &  0
\end{pmatrix}
,
\end{equation*}
where $( \,\ \vert \,\ )$  is the super-invariant super-symmetric bilinear 
form.  Choose $X_{\alpha} \in \jj_{\alpha}$, for each root $\alpha$, 
satisfying the conditions (2.1), (2.2) and $c_{\alpha_{1}, \alpha_{2}} =1$, 
where $H_{\alpha_{i}} = \alpha^{\vee}_{i}$.  Then the fields
{\allowdisplaybreaks %
\begin{align*}
G'(z) &:= X_{\alpha_{1}}(z)\psi^{\alpha_{1}}(z)
+X_{-\alpha_{2}}(z)\psi^{-\alpha_{2}}(z),   \\
G^{\prime \ast}(z) &:= X_{-\alpha_{1}}(z)\psi^{-\alpha_{1}}(z)
+X_{\alpha_{2}}(z)\psi^{\alpha_{2}}(z),  \\
A(z) &:= K \left\{-:\psi^{\alpha_{1}}(z)\psi_{\alpha_{1}}(z):
+:\psi^{\alpha_{2}}(z)\psi_{\alpha_{2}}(z):\right\} 
+ (H_{\alpha_{1}}- H_{\alpha_{2}})(z),  \\
B(z) &:= K \left\{-:\pa \psi^{\alpha_{1}}(z)\psi_{\alpha_{1}}(z):
+:\psi^{\alpha_{2}}(z) \pa \psi_{\alpha_{2}}(z):\right\}   \\
&+ \sum_{i=1,2} \left\{ :X_{\alpha_{i}}(z)X_{-\alpha_{i}}(z):
+ :H_{\alpha_{i}}(z) \psi^{\alpha_{i}}(z) \psi_{\alpha_{i}}(z): \right\}  \\
&- \pa H_{\alpha_2}(z) 
+ :X_{\alpha_{1}+\alpha_{2}}(z) \psi^{\alpha_{1}}(z) \psi^{\alpha_{2}}(z):
- :X_{-\alpha_{1}-\alpha_{2}}(z) \psi_{\alpha_{1}}(z) \psi_{\alpha_{2}}(z):
\end{align*}
}
satisfy the following product formulas:
{\allowdisplaybreaks %
\begin{align*}
G'(z)G'(w) &= G^{\prime \ast}(z)G^{\prime \ast}(w) = 0,  \\
G'(z)G^{\prime \ast}(w) &= \frac{-2K}{(z-w)^3} + \frac{A(w)}{(z-w)^2}
+ \frac{B(w)}{z-w},  \\
A(z)G'(w) &= \frac{K+1}{z-w}G'(w),  \\
A(z)G^{\prime \ast}(w) &= \frac{-(K+1)}{z-w}G^{\prime \ast}(w),   \\
B(z)G'(w) &= (K+1) \left\{\frac{G'(w)}{(z-w)^2} + \frac{\pa G'(w)}
{z-w} \right\}, \\
B(z)G^{\prime \ast}(w) &= (K+1) \left\{\frac{2G^{\prime \ast}(w)}{(z-w)^2} 
+ \frac{\pa G^{\prime \ast}(w)}{z-w} \right\},  \\
A(z)A(w)  &= \frac{-4K(K+1)}{(z-w)^2},   \\
A(z)B(w)  &= (K+1) \left\{ \frac{-2K}{(z-w)^3} +\frac{A(w)}
{(z-w)^2} \right\},   \\
B(z)A(w)  &= (K+1) \left\{ \frac{2K}{(z-w)^3} +\frac{A(w)}{(z-w)^2} 
+\frac{\pa A(w)}{z-w} \right\},   \\
B(z)B(w)  &= (K+1) \left\{ \frac{2B(w)}{(z-w)^2} +\frac{\pa B(w)}
{z-w} \right\}.
\end{align*}
}
\end{thm}

Putting  
{\allowdisplaybreaks %
\begin{align*}
L(z) &:= \frac{1}{K+1}\left\{B(z)-\frac{1}{2} \pa A(z)\right\} ,   \qquad
J(z) := \frac{1}{K+1}A(z),   \\
G(z) &:= \frac{1}{K+1}G'(z),  \qquad \qquad \qquad \qquad
G^{\ast}(z) := G^{\prime \ast}(z),  \\
K' &:= \frac{-2K}{K+1},
\end{align*}
}
one can rewrite the above into the the standard product formulas of the N=2 
superconformal algebra as follows: 

\begin{thm}{Corollary 2.1}
{\allowdisplaybreaks %
\begin{align*}
L(z)L(w) &= \frac{\frac{1}{2}K'}{(z-w)^4} + \frac{2L(w)}{(z-w)^2} 
+ \frac{\pa L(w)}{z-w}, \\
L(z)J(w) &= \frac{J(w)}{(z-w)^2} +\frac{\pa J(w)}{z-w},  \qquad
J(z)L(w) = \frac{J(w)}{(z-w)^2},   \\
L(z)G(w) &= \frac{\frac{3}{2}G(w)}{(z-w)^2} +\frac{\pa G(w)}{z-w},  \quad
G(z)L(w) = \frac{\frac{3}{2}G(w)}{(z-w)^2} +\frac{\frac{1}{2} \pa G(w)}
{z-w},  \\
L(z)G^{\ast}(w) &= \frac{\frac{3}{2}G^{\ast}(w)}{(z-w)^2} 
+\frac{\pa G^{\ast}(w)}{z-w},  \quad
G^{\ast}(z)L(w) = \frac{\frac{3}{2}G^{\ast}(w)}{(z-w)^2} 
+\frac{\frac{1}{2} \pa G^{\ast}(w)}{z-w}, \\ 
J(z)J(w) &= \frac{2K'}{(z-w)^2}, \\
J(z)G(w) &= \frac{G(w)}{z-w}, \qquad
G(z)J(w) = \frac{-G(w)}{z-w}, \\
J(z)G^{\ast}(w) &= \frac{-G^{\ast}(w)}{z-w}, \qquad
G^{\ast}(z)J(w) = \frac{G^{\ast}(w)}{z-w}, \\
G(z)G^{\ast}(w) &= \frac{K'}{(z-w)^3} +\frac{J(w)}{(z-w)^2} 
+\frac{L(w) + \frac{1}{2} \pa J(w)}{z-w},  \\
G^{\ast}(z)G(w) &= \frac{K'}{(z-w)^3} - \frac{J(w)}{(z-w)^2} 
+\frac{L(w) - \frac{1}{2} \pa J(w)}{z-w},  \\
G(z)G(w) &= G^{\ast}(z)G^{\ast}(w) =0.
\end{align*}
}
\end{thm}

We note that this superconformal algebra $\aaa$ has important commutants.
The following theorem is easily shown from (1.5) and (1.6b):

\begin{thm}{Theorem 2.2}
Introduce two fields:
$$d_1(z) := X_{\alpha_1 + \alpha_2}(z) + :\psi_{\alpha_1}(z) 
\psi_{\alpha_2}(z):$$
and
$$d_2(z) := X_{-\alpha_1 - \alpha_2}(z) - :\psi^{\alpha_1}(z) 
\psi^{\alpha_2}(z):.$$
Then,
$$d_i(z) G(w) = d_i(z) G^{\ast}(w) =0 \,\ \text{for} \,\ i=1,2.$$
\end{thm}

This implies, by the Borcherds-Jacobi identity (2.9), that
\begin{equation*}
 [d_{i(m)}, G_{(n)}] = [d_{i(m)}, G^{\ast}_{(n)}] = 0     \tag{2.10}
\end{equation*}
for all $m,n \in \zzz$ and $i=1,2$.

Let $V$ be the tensor product of a highest weight 
$\widehat{sl}(2 \vert 1)$-module and the symmetric algebra over 
$ \{ \psi^{\alpha_i}(n); \,\ i=1,2 \,\ \text{and} \,\ n <0 \}$.
Then $V$ is an $\aaa$-module and, by (2.10), all simultaneous eigenspaces 
of $d_{i(m)}$, $i=1,2$ and $m \in \zzz$, are stable under the action of 
$G_{(n)}$ and $G^{\ast}_{(n)}$, and so are $\aaa$-submodules.

\section*{3. The centerless N=4 superconformal algebra via 
$\widehat{A}(1, 1)$}

Let us consider the superalgebra $sl(2 \vert 2)$  with simple roots 
$\Pi = \{ \alpha_{1},\alpha_{2},\alpha_{3} \}$  and  Chevalley generators 
$(e_{i}, f_{i}, \alpha^{\vee}_{i})_{i=1,2,3}$,  satisfying
\begin{equation*}
     \left( \langle \alpha^{\vee}_{i}, 
\alpha_{j}\rangle \right)_{i, j =1,2,3} =
\begin{pmatrix}
    0  &   1  &  0   \\
    1  &  -2  &  1  \\
    0  &   1  &  0
\end{pmatrix}
\tag{3.1}.
\end{equation*}
The algebra $sl(2 \vert 2)$ has a super-invariant super-symmetric bilinear 
form  $(\,\ \vert \,\ )$  such that
\begin{equation*}
     \left( (\alpha^{\vee}_{i} \vert \alpha^{\vee}_{j}) \right)_{i, j =1,2,3}
   =
\begin{pmatrix}
    0  &   1  &  0   \\
    1  &  -2  &  1  \\
    0  &   1  &  0
\end{pmatrix}
\tag{3.2}.
\end{equation*}
  From this, one sees that  $sl(2 \vert 2)$ is not simple, since the Cartan 
matrix (3.1) is singular of rank 2, or in other words, the inner product 
defined by (3.2) is degenerate, and that 
 $\alpha^{\vee}_{1} - \alpha^{\vee}_{3}$  
spans the center.  The quotient simple superalgebra 
$sl(2 \vert 2)/ \ccc \cdot (\alpha^{\vee}_{1} - \alpha^{\vee}_{3})$ 
is called  $A(1, 1)$.  Its Cartan subalgebra $\hhh$ is, therefore, 
2-dimensional, and each $\alpha_{i}$ naturally defines a linear form on 
$\hhh$,  which is also denoted by $\alpha_{i}$ using the same characters.
Under this notation, the induced inner product on the dual space 
$\hhh^{\ast}$ satisfies 
\begin{equation*}
     \left( (\alpha_{i} \vert \alpha_{j}) \right)_{i, j =1,2,3} =
\begin{pmatrix}
    0  &   1  &  0   \\
    1  &  -2  &  1  \\
    0  &   1  &  0
\end{pmatrix}
\tag{3.3}.
\end{equation*}
For simplicity, we write 
$(m_{1}, m_{2}, m_{3})$  in place of $\sum^{3}_{i=1} m_{i} \alpha_{i}$,
and also
$-(m_{1}, m_{2}, m_{3})$  in place of $-\sum^{3}_{i=1} m_{i} \alpha_{i}$.
Then the sets $\Delta^{+}$, $\Delta^{+}_{even}$  and $\Delta^{+}_{odd}$ 
of roots of $A(1, 1)$ are given as follows:
\begin{align*}
\Delta^{+}_{even} &= \{ (0,1,0), \,\ (1,1,1) \},   \\
\Delta^{+}_{odd} &= \{ (1,0,0), \,\ (0,0,1), \,\ (1,1,0), \,\ (0,1,1) \}, \\
\Delta^{+} &= \Delta^{+}_{even} \cup \Delta^{+}_{odd}. 
\end{align*}
We note that the square length of a root $\alpha$ is equal to $-2$ if 
$\alpha = \pm \alpha_2$,  $2$  if $\alpha = \pm (1,1,1)$, and $0$ if 
$\alpha$ is odd. 
For each root $\alpha=(m_{1}, m_{2}, m_{3})$ we choose a root vecter 
$X_{\alpha}$ =  $X_{(m_{1}, m_{2}, m_{3})}$   in such a way that
\begin{align*}
e_{(1,1,0)} &:= [e_1, e_2], \quad e_{(0,1,1)} := [e_2, e_3],  \quad
e_{(1,1,1)} := [e_1, e_{(0,1,1)}]= [e_{(1,1,0)}, e_3],   \\
f_{(1,1,0)} &:= [f_1, f_2], \quad f_{(0,1,1)} := [f_2, f_3],  \quad
f_{(1,1,1)} := [f_1, f_{(0,1,1)}]= [f_{(1,1,0)}, f_3].
\end{align*}

We now consider the affinization $\widehat{A}(1, 1)$, and the following 
fields:
{\allowdisplaybreaks %
\begin{align*}
G_{1}(z) &:= X_{(1,1,0)}(z)\psi^{\alpha_{1}}(z)
-X_{-(0,0,1)}(z)\psi^{-\alpha_{3}}(z),  \\
G^{\ast}_{1}(z) &:= 
-X_{-(1,1,0)}(z)\psi^{-\alpha_{1}}(z) +X_{(0,0,1)}(z)\psi^{\alpha_{3}}(z), \\
G_{2}(z) &:= X_{(1,0,0)}(z)\psi^{\alpha_{1}}(z) 
- X_{-(0,1,1)}(z)\psi^{-\alpha_{3}}(z),  \\
G^{\ast}_{2}(z) &:= 
-X_{-(1,0,0)}(z)\psi^{-\alpha_{1}}(z) +X_{(0,1,1)}(z)\psi^{\alpha_{3}}(z), \\
A_{1}(z) &:= K \left\{:\psi^{\alpha_{3}}(z)\psi_{\alpha_{3}}(z):
-:\psi^{\alpha_{1}}(z)\psi_{\alpha_{1}}(z):\right\} 
- \alpha^{\vee}_{2}(z), \\
A_{2}(z) &:= K \left\{:\psi^{\alpha_{3}}(z)\psi_{\alpha_{3}}(z):
-:\psi^{\alpha_{1}}(z)\psi_{\alpha_{1}}(z):\right\} 
+ \alpha^{\vee}_{2}(z), \\
B_{1}(z) &:= K \left\{:\psi^{\alpha_{3}}(z) \pa \psi_{\alpha_{3}}(z):
-:\pa \psi^{\alpha_{1}}(z)\psi_{\alpha_{1}}(z):\right\} \\
&-:X_{(1,1,0)}(z)X_{-(1,1,0)}(z):+:X_{(0,0,1)}(z)X_{-(0,0,1)}(z):  \\
&-:(\alpha^{\vee}_{1}+\alpha^{\vee}_{2})(z)\psi^{\alpha_{1}}(z)
\psi_{\alpha_{1}}(z):
-:\alpha^{\vee}_{3}(z)\psi^{\alpha_{3}}(z)\psi_{\alpha_{3}}(z):   \\
&+ \pa \alpha^{\vee}_{3}(z)
+:X_{(1,1,1)}(z)\psi^{\alpha_{1}}(z)\psi^{\alpha_{3}}(z):
+:X_{-(1,1,1)}(z)\psi_{\alpha_{1}}(z)\psi_{\alpha_{3}}(z):,  \\
B_{2}(z) &:= K \left\{:\psi^{\alpha_{3}}(z) \pa \psi_{\alpha_{3}}(z):
-:\pa \psi^{\alpha_{1}}(z)\psi_{\alpha_{1}}(z):\right\} \\
&-:X_{(1,0,0)}(z)X_{-(1,0,0)}(z):-:X_{(0,1,1)}(z)X_{-(0,1,1)}(z):   \\
&-:(\alpha^{\vee}_{2}+\alpha^{\vee}_{3})(z)\psi^{\alpha_{3}}(z)
\psi_{\alpha_{3}}(z):
-:\alpha^{\vee}_{1}(z)\psi^{\alpha_{1}}(z)\psi_{\alpha_{1}}(z):   \\
&+ \pa (\alpha^{\vee}_{2} + \alpha^{\vee}_{3})(z)
+:X_{(1,1,1)}(z)\psi^{\alpha_{1}}(z)\psi^{\alpha_{3}}(z):  \\
&+:X_{-(1,1,1)}(z)\psi_{\alpha_{1}}(z)\psi_{\alpha_{3}}(z):.
\end{align*}
}

Then Theorem 2.1 applied to each quadruplet 
$(G_{i}, G^{\ast}_{i}, A_{i}, B_{i})$, $i = 1,2$, gives the following:
\begin{equation*}
G_{i}(z)G_{i}(w) = G^{\ast}_{i}(z)G^{\ast}_{i}(w) = 0,      \tag{3.4a}
\end{equation*}
\begin{equation*}
G_{i}(z)G^{\ast}_{i}(w) = \frac{-2K}{(z-w)^3} + \frac{A_{i}(w)}{(z-w)^2}
+ \frac{B_{i}(w)}{z-w},    \tag{3.4b}
\end{equation*}
\begin{equation*}
A_{i}(z)G_{i}(w) = \frac{K+1}{z-w}G_{i}(w),    \tag{3.5a}
\end{equation*}
\begin{equation*}
A_{i}(z)G^{\ast}_{i}(w) = \frac{-(K+1)}{z-w}G^{\ast}_{i}(w),    \tag{3.5b}
\end{equation*}
\begin{equation*}
B_{i}(z)G_{i}(w) = (K+1) \left\{\frac{G_{i}(w)}{(z-w)^2} + 
\frac{\pa G_{i}(w)}{z-w} \right\},    \tag{3.6a}
\end{equation*}
\begin{equation*}
B_{i}(z)G^{\ast}_{i}(w) = (K+1) \left\{\frac{2G^{\ast}_{i}(w)}{(z-w)^2} + 
\frac{\pa G^{\ast}_{i}(w)}{z-w} \right\}.    \tag{3.6b}
\end{equation*}

One also has the following:
\begin{equation*}
e_2(z)G_1(w) = 0,          \qquad    
e_2(z)G_2(w) = \frac{-G_1(w)}{z-w},  \tag{3.7a} 
\end{equation*}
\begin{equation*}
e_2(z)G^{\ast}_{1}(w) = \frac{G^{\ast}_2(w)}{z-w},  \qquad
e_2(z)G^{\ast}_{2}(w) = 0,        \tag{3.7b}   
\end{equation*}
\begin{equation*}
f_2(z)G_1(w) = \frac{G_2(w)}{z-w},   \qquad
f_2(z)G_2(w) = 0,                    \tag{3.8a}
\end{equation*}
\begin{equation*}
f_2(z)G^{\ast}_{1}(w) = 0,           \qquad  \qquad
f_2(z)G^{\ast}_{2}(w) = \frac{-G^{\ast}_1(w)}{z-w}.     \tag{3.8b}
\end{equation*}
\begin{equation*}
\alpha^{\vee}_2(z)G_{1}(w) = \frac{-G_1(w)}{z-w},      \qquad
\alpha^{\vee}_2(z)G_{2}(w) = \frac{G_2(w)}{z-w}.     \tag{3.9a}
\end{equation*}
\begin{equation*}
\alpha^{\vee}_2(z)G^{\ast}_{1}(w) = \frac{G^{\ast}_1(w)}{z-w},  \qquad
\alpha^{\vee}_2(z)G^{\ast}_{2}(w) = \frac{-G^{\ast}_2(w)}{z-w},  \tag{3.9b}
\end{equation*}
and
\begin{equation*}
A_1(z)G_2(w)=\frac{K-1}{z-w}G_2(w),  \qquad
A_1(z)G^{\ast}_2(w)=\frac{1-K}{z-w}G^{\ast}_2(w),    \tag{3.10a}
\end{equation*}
\begin{equation*}
A_2(z)G_1(w)=\frac{K-1}{z-w}G_1(w),  \qquad
A_2(z)G^{\ast}_1(w)=\frac{1-K}{z-w}G^{\ast}_1(w).    \tag{3.10b}
\end{equation*}

The operator products  $G_{i}(z)G^{\ast}_{j}(w)$ for $i \ne j$ are 
obtained from  Lemma 1.1 as follows:
\begin{equation*}
G_1(z)G_2(w)  = G^{\ast}_1(z)G^{\ast}_2(w) =0,    \tag{3.11a}
\end{equation*}
\begin{equation*}
\begin{split}
G_1(z)G^{\ast}_2(w)  &= -\frac{2e_2(w)}{(z-w)^2}-\frac{\pa e_2(w)}{z-w}  \\
&+\frac{1}{z-w} \{e_2(w) \left( :\psi^{\alpha_3}(w)\psi_{\alpha_3}(w):
-:\psi^{\alpha_1}(w)\psi_{\alpha_1}(w): \right)  \\
&- :X_{(1,1,0)}(w)X_{-(1,0,0)}(w):- :X_{(0,1,1)}(w)X_{-(0,0,1)}(w):  \}, \\
\end{split}              \tag{3.11b}
\end{equation*}
\begin{equation*}
\begin{split}
G_2(z)G^{\ast}_1(w)  &= \frac{2f_2(w)}{(z-w)^2}+\frac{\pa f_2(w)}{z-w}  \\
&+\frac{1}{z-w} \{f_2(w) \left( :\psi^{\alpha_1}(w)\psi_{\alpha_1}(w):
-:\psi^{\alpha_3}(w)\psi_{\alpha_3}(w): \right)   \\
&- :X_{(1,0,0)}(w)X_{-(1,1,0)}(w):- :X_{(0,0,1)}(w)X_{-(0,1,1)}(w):  \}.
\end{split}   \tag{3.11c}
\end{equation*}

We now introduce the field
\begin{equation*}
\begin{split}
D(z) &:= B_{1}(z)-B_{2}(z)+ \pa \alpha^{\vee}_{2}(z)  \\
&= :X_{(1,0,0)}(z)X_{-(1,0,0)}(z):+ :X_{(0,1,1)}(z)X_{-(0,1,1)}(z):   \\
&- :X_{(1,1,0)}(z)X_{-(1,1,0)}(z):- :X_{(0,0,1)}(z)X_{-(0,0,1)}(z):   \\
&+\alpha^{\vee}_{2}(z) \left(:\psi^{\alpha_{3}}(z) \psi_{\alpha_{3}}(z):
-:\psi^{\alpha_{1}}(z) \psi_{\alpha_{1}}(z):  \right).
\end{split}     \tag{3.12}
\end{equation*}

Then one has
\begin{equation*}
\begin{split}
D(z)e_2(w) &= \frac{2}{z-w} 
\{ :X_{(1,1,0)}(w)X_{-(1,0,0)}(w):+ :X_{(0,1,1)}(w)X_{-(0,0,1)}(w):  \\
&+e_2(w) \left( :\psi^{\alpha_1}(w)\psi_{\alpha_1}(w):
-:\psi^{\alpha_3}(w)\psi_{\alpha_3}(w): \right) \},  
\end{split}   \tag{3.13a}
\end{equation*}
\begin{equation*}
\begin{split}
D(z)f_2(w) &= \frac{2}{z-w} \{
 :X_{(1,0,0)}(w)X_{-(1,1,0)}(w):+ :X_{(0,0,1)}(w)X_{-(0,1,1)}(w):   \\
&+f_2(w) \left( :\psi^{\alpha_3}(w)\psi_{\alpha_3}(w):
-:\psi^{\alpha_1}(w)\psi_{\alpha_1}(w): \right)  \} .
\end{split}     \tag{3.13b}
\end{equation*}

Let $\tilde{\aaa}$ be the  vertex algebra generated by $G_i$, $G^{\ast}_i$ 
($i = 1, 2$),   $e_2$, $f_2$, $\alpha^{\vee}_2$, and  $\iii$ its ideal 
generated by $D$, and we consider the vertex algebra  
$\aaa := \tilde{\aaa}/\iii$.  Then, by (3.13a) and (3.13b), the third 
terms in the right-hand sides of (3.11b) and (3.11c) vanish in $\aaa$, 
and one has
\begin{equation*}
G_1(z)G^{\ast}_2(w)  = - \left\{ \frac{2e_2(w)}{(z-w)^2}+\frac{\pa e_2(w)}
{z-w} \right\} , \tag{3.11'b}
\end{equation*}
\begin{equation*}
G_2(z)G^{\ast}_1(w)  = \frac{2f_2(w)}{(z-w)^2}+\frac{\pa f_2(w)}{z-w}.
 \tag{3.11'c}
\end{equation*}

As a cost of this ideal $\iii$, we have to assume henceforward that $K=0$, 
since otherwise $G_i$ and $G^{\ast}_i$ belong to  $\iii$ and the vertex 
algebra $\aaa$  collapses.  Then
\begin{equation*}
A_1 = -A_2 = -\alpha^{\vee}_2,    \tag{3.14}
\end{equation*}
and all other products are computed from (3.4) $\sim$ (3.11') by the 
Borcherds-Jacobi identity (2.9) in a similar way as is explained 
in Section 2.  And by putting
{\allowdisplaybreaks %
\begin{align*}
L(z) &:= B_1(z)-\frac{1}{2} \pa A_1(z) = B_2(z)-\frac{1}{2} \pa A_2(z), \\
h(z) &:= - \alpha^{\vee}_2(z),   \qquad
e(z) := -e_2(z),  \qquad
f(z) := f_2(z),
\end{align*}
}
one sees that these fields satisfy the product formulas of the N=4 
superconformal algebra:

\begin{thm}{Theorem 3.1}
{\allowdisplaybreaks %
\begin{align*}
L(z)L(w) &= \frac{2L(w)}{(z-w)^2} + \frac{\pa L(w)}{z-w}, \\
L(z)X(w) &= \frac{X(w)}{(z-w)^2} +\frac{\pa X(w)}{z-w},  \qquad
X(z)L(w) = \frac{X(w)}{(z-w)^2}, \\
X(z)Y(w) &= \frac{[X, Y](w)}{z-w}  \qquad \qquad \qquad 
\text{for} \,\ X,Y = h, \,\ e, \,\ \text{or} \,\ f, \\
L(z)F(w) &= \frac{\frac{3}{2}F(w)}{(z-w)^2} +\frac{\pa F(w)}{z-w},  \\
F(z)L(w) &= \frac{\frac{3}{2}F(w)}{(z-w)^2} +\frac{\frac{1}{2} \pa F(w)}
{z-w} \qquad \qquad \qquad \text{for} \,\ F=G_i, \,\ G^{\ast}_i,  \\
h(z)G_1(w) &=-G_1(z)h(w)= \frac{G_1(w)}{z-w}, \quad
h(z)G^{\ast}_1(w) =-G^{\ast}_1(z)h(w)= \frac{-G^{\ast}_1(w)}{z-w}, \\
h(z)G_2(w) &=-G_2(z)h(w)= \frac{-G_2(w)}{z-w}, \quad
h(z)G^{\ast}_2(w) =-G^{\ast}_2(z)h(w)= \frac{G^{\ast}_2(w)}{z-w}, \\
e(z)G_1(w) &=G_1(z)e(w)= 0, \qquad  \qquad 
e(z)G^{\ast}_1(w) =-G^{\ast}_1(z)e(w)= \frac{-G^{\ast}_2(w)}{z-w}, \\
e(z)G_2(w) &=-G_2(z)e(w)= \frac{G_1(w)}{z-w}, \qquad
e(z)G^{\ast}_2(w) =G^{\ast}_2(z)e(w)= 0, \\
f(z)G_1(w) &=-G_1(z)f(w)= \frac{G_2(w)}{z-w}, \qquad
f(z)G^{\ast}_1(w) = G^{\ast}_1(z)f(w)= 0, \\
f(z)G_2(w) &=G_2(z)f(w)= 0, \qquad
f(z)G^{\ast}_2(w) =-G^{\ast}_2(z)f(w)= \frac{-G^{\ast}_1(w)}{z-w},  \\
G_i(z)G_j(w) &= G^{\ast}_i(z)G^{\ast}_j(w) =0,      \\
G_1(z)G^{\ast}_1(w) &= G^{\ast}_2(z)G_2(w) = 
\frac{h(w)}{(z-w)^2} +\frac{L(w) + \frac{1}{2} \pa h(w)}{z-w},  \\
G_2(z)G^{\ast}_2(w) &= G^{\ast}_1(z)G_1(w) = 
\frac{-h(w)}{(z-w)^2} +\frac{L(w) - \frac{1}{2} \pa h(w)}{z-w},  \\
G_1(z)G^{\ast}_2(w) &= -G^{\ast}_2(z)G_1(w) = 
\frac{2e(w)}{(z-w)^2} +\frac{ \pa e(w)}{z-w},  \\
G_2(z)G^{\ast}_1(w) &= -G^{\ast}_1(z)G_2(w) = 
\frac{2f(w)}{(z-w)^2} +\frac{ \pa f(w)}{z-w}.
\end{align*}
}
\end{thm}

Commutants for $\aaa$ are given as follows in this case:

\begin{thm}{Theorem 3.2}
Consider two fields:
$$d_1(z) := X_{(1,1,1)}(z) + : \psi_{\alpha_1}(z) \psi_{\alpha_3}(z): $$
and
$$d_2(z) := X_{-(1,1,1)}(z) + : \psi^{\alpha_1}(z) \psi^{\alpha_3}(z):.$$
Then  $$d_i(z) G_j(w) = d_i(z) G^{\ast}_j(w)=0 \,\ \text{and} \,\  \,\
d_i(z) X(w) =0,$$
for $i,j=1,2$  and $X= h,e,f$.
\end{thm}

  From this theorem, one has
$$ [d_{i(m)}, G_{j(n)}] = [d_{i(m)}, G^{\ast}_{j(n)}] =  0$$
and 
$$ [d_{i(m)}, X_{(n)}] = 0, $$
for $i,j = 1,2$,  $X=h,e,f$, and $m,n \in \zzz$.
Then, by (3.12), all $d_{i(m)}$'s commute with $D_{(n)}$ and so, when we 
consider representations, commute with the action of the vertex algebra 
$\aaa$.  And all simultaneous eigenspaces of $d_{i(m)}$, $i=1,2$ and  
$m \in \zzz$, are $\aaa$-submodules.


\begin{thebibliography}{[1]}

\bibitem{HT}  S. Hosono and A. Tsuchiya : Lie algebra cohomology and N=2 
SCFT based on the GKO construction,  Commun. Math. Phys. 136 (1991), 451-486.

\bibitem{K1}  V. G. Kac : Lie superalgebras,  Adv. in Math. 26 (1977), 
8-96.

\bibitem{K2}  V. G. Kac : Infinite dimensional Lie algebras, third 
edition, Cambridge University Press, 1990.

\bibitem{K3}  V. G. Kac : Vertex Algebras for Beginners, University 
Lecture Series Vol. 10, American Mathematical Society, 1996.

\bibitem{K4}  V. G. Kac : Superconformal algebras and transitive group 
actions on quadrics, preprint, 1996.

\end{thebibliography}
\end{document}